\documentclass[iop]{emulateapj}

\shorttitle{Supergranulation spectrum}
\shortauthors{Lord, Cameron, Rast, Rempel and Roudier}

\begin{document}

\title{The Role of Subsurface Flows in Solar Surface Convection: Modeling the Spectrum of Supergranular and Larger Scale Flows}

\author{J. W. Lord}
\affil{Department of Astrophysical and Planetary Sciences, Laboratory for Atmospheric and Space Physics, University of Colorado, Boulder, CO 80309}

\author{R. H. Cameron}
\affil{Max-Planck-Institut f\"ur Sonnensystemforschung, Justus-von-Liebig-Weg 3, 37077 G\"ottingen, Germany}

\author{M. P. Rast\altaffilmark{1}}
\affil{Department of Astrophysical and Planetary Science, Laboratory for Atmospheric and Space Physics, University of Colorado, Boulder, CO 80309}

\author{M. Rempel}
\affil{High Altitude Observatory, National Center for Atmospheric Research\altaffilmark{2}, Boulder, CO 80307}

\and

\author{T. Roudier}
\affil{Institut de Recherche en Astrophysique et Plan\'etologie, Universit\'e de Toulouse, Centre national de la recherche scientifique (CNRS), 31400 Toulouse, France}

\altaffiltext{1}{Corresponding author mark.rast@lasp.colorado.edu}
\altaffiltext{2}{NCAR is sponsored by the National Science Foundation}

\begin{abstract}

We model the solar horizontal velocity power spectrum at scales larger than granulation using a two-component approximation to the mass continuity equation. The model takes four times the density scale height as the integral (driving) scale of the vertical motions at each depth. Scales larger than this decay with height from the deeper layers. Those smaller are assumed to follow a Kolomogorov turbulent cascade, with the total power in the vertical convective motions matching that required to transport the solar luminosity in a mixing length formulation. These model components are validated using large scale radiative hydrodynamic simulations. We reach two primary conclusions: 1. The model predicts significantly more power at low wavenumbers than is observed in the solar photospheric horizontal velocity spectrum. 2. Ionization plays a minor role in shaping the observed solar velocity spectrum by reducing convective amplitudes in the regions of partial helium ionization. The excess low wavenumber power is also seen in the fully nonlinear three-dimensional radiative hydrodynamic simulations employing a realistic equation of state. This adds to other recent evidence suggesting that the amplitudes of large scale convective motions in the Sun are significantly lower than expected. Employing the same feature tracking algorithm used with observational data on the simulation output, we show that the observed low wavenumber power can be reproduced in hydrodynamic models if the amplitudes of large scale modes in the deep layers are artificially reduced. Since the large scale modes have reduced amplitudes, modes on the scale of supergranulation and smaller remain important to convective heat flux even in the deep layers, suggesting that small scale convective correlations are maintained through the bulk of the solar convection zone.

\end{abstract}

\keywords{Convection --- Turbulence --- Sun: photosphere --- Sun: granulation --- Sun: interior}

\section{Introduction}

Solar supergranulation is observed as horizontal divergent flow within magnetic network boundaries~\citep{lei62,sim64}, either by Doppler imaging away from disk center~\citep{hat00} or by correlation~\citep{nov88,der04,meu07} or structure tracking near disk center~\citep{rou99,rou12}. The power spectrum of the horizontal motions shows a characteristic peak at horizontal scales ranging from approximately 20Mm to 50Mm, and the motions at these scales are identified as supergranulation. There is a dramatic drop in spectral power for scales larger than supergranulation with very weak giant cell flows only recently confirmed by observations~\citep{hat13}.

The physical origin of the supergranular length scale remains a mystery. Suggestions range from possible dynamical effects of the second ionization of Helium~\citep{lei62,sim64,nov81} to spatial correlation or self organization of granular flows~\citep{rie00,ras03,cro07}. Radiative hydrodynamic simulations of solar surface convection fail to yield clear evidence for supergranulation, even in very large domains spanning up to $96\mbox{Mm}$ by $96\mbox{Mm}$ in width and $20$Mm in depth~\citep{ste09,ust10}. Recent simulations in even larger domains of up to $196\times196\times49\ \mbox{Mm}^3$ suggest that the domain depth, and the consequent stratification captured by the simulation, may be as critical as domain width~\citep{lor14}. Based on these broad and deep simulations of solar surface convection we have developed a model of the convective velocity spectrum which reproduces the simulation spectrum and provides insight into how the deep convective flows help build the observed photospheric spectrum.

The model assumes that, at each depth, vertical motions are driven at scales four times the local density scale height. The amplitudes of smaller scale motions is taken to be consistent with the spectrum of unstratified and incompressible turbulence. Larger scale vertical motions imprint from below with reduced amplitude and are observed as primarily horizontal flows at the surface~\citep{spr90}. In other words, modes with wavelengths smaller than the integral (driving) scale are assumed to have amplitudes that follow the spectrum of isotropic homogeneous turbulence given by~\citet{kol41}, while vertical motions of scales larger than the integral scale are assumed to decay with height from their driving depth. The integrated power of the vertical velocity is determined using a mixing length model of energy transport, and the spectrum of horizontal velocity follows from the vertical velocity spectrum at each depth by mass continuity.

Key scalings in the model are verified using the radiative hydrodynamic simulations of~\cite{lor14}, and the simplified model spectra agree with those of the simulations over a wide range of wavenumbers. They also match observations at supergranular scales. However, power at lower wavenumbers, in both the model and radiative hydrodynamic simulation spectra, significantly exceeds that observed. This suggests either that large scale flows deep in the solar convection zone are weaker than predicted by convection simulations or that rotation and the consequent near surface shear layer, not captured in our studies, plays a decisive role in masking large scale motions. We note however, that recent helioseismic observations~\citep{han10,han12} and global scale numerical simulations, with and without a near surface shear layer~\citep{hot14}, also suggest that large scale convection in the Sun is weaker than numerical models predict. It is possible that magnetic fields play a role, that convection in the Sun is fundamentally magnetized. Preliminary studies of radiative magnetohydrodynamic simulations~\citep{lor14} in very large domains show some suppression of low wavenumber power in highly magnetized solutions, though the mechanism is still under investigation and the effect so far appears insufficient to explain solar observations. In this paper we focus on strictly hydrodynamic effects to elucidate the important role of stratification and the secondary influence of ionization in shaping the photospheric horizontal velocity power spectrum at supergranular and larger scales.

In \S{2} we describe the simplified two-component continuity balance on which our model is based, and transform the balance equations into relationships between the vertical and horizontal velocity spectra. We use these relationships to identify the driving scale of the modes and demonstrate that these relationships hold in fully compressible hydrodynamic simulations. In \S{3} we describe the construction of the mixing length atmosphere which sets the amplitude of the model spectrum, identify two possible decay rates for the large scale modes, and explicitly outline the model steps employed in the construction of the surface horizontal velocity spectrum. In \S{4} we test the components of the model spectrum against the full radiative hydrodynamic solutions and verify that the model can reproduce the shape of the spectrum produced by those simulations. In \S{5} we discuss the results of the model spectrum, focusing on the spectrum at supergranular scales and larger. We show, using feature tracking, that for scales larger than supergranulation the radiative hydrodynamic spectra can only match the observations when the convective forcing is removed entirely below 10 Mm. We conclude, in \S{6} with a discussion of the broader implications of the weak low wavenumber amplitudes to our understanding of deep solar convection.

\section{Mass continuity and the effects of stratification}

We use the equation of mass continuity to examine how stratification affects flow velocity. Explicitly,

\begin{equation}
\frac{\partial\rho}{\partial t}+ \nabla \cdot (\rho \bold{u})=\frac{\partial\rho}{\partial t}+\rho(\frac{\partial \mbox{u}_x}{\partial x}+\frac{\partial \mbox{u}_y}{\partial y}+\frac{\partial \mbox{u}_z}{\partial z})+\mbox{u}_x\frac{\partial \rho}{\partial x}+\mbox{u}_y\frac{\partial \rho}{\partial y}+\mbox{u}_z\frac{\partial \rho}{\partial z}=0
\end{equation}
\noindent
where $\rho$ is the mass density, $\bold{u}$ is the fluid velocity, and subscripts $x$ and $y$ and $z$ indicate components in Cartesian coordinates. We ignore curvature, and take gravity, and thus increasing density in the stratified domain, to be in the positive $z$ direction.

In the solar convection zone we can make a number of further simplifying assumptions. Since we are looking for the statistically steady velocity amplitudes over time periods much longer than the convective turnover time, we take $\frac{\partial\rho}{\partial t}\rightarrow0$. Moreover, we know from hydrodynamic simulations that the horizontal gradients of the density are small compared to the vertical stratification below the first few hundred kilometers beneath the solar photosphere, so $\mbox{u}_x\frac{\partial \rho}{\partial x}$ and $\mbox{u}_y\frac{\partial \rho}{\partial y}$ are ignored. Together these assumptions yield an anelastic-like continuity equation~\citep{gou69} that maintains the steady state stratification by balancing the vertical advection of mass with the divergence of the flow,
\begin{equation}
\label{conteqn}
{\bf{\nabla}}_h\cdot{\bf{u}}_h=-\frac{\partial \mbox{u}_z}{\partial z}-\frac{\mbox{u}_z}{H_{\rho}}\ ,
\end{equation}
where ${\bf{u}}_h=\mbox{u}_x\hat i+\mbox{u}_y\hat j$, ${\bf{\nabla}}_h=\hat i\frac{\partial}{\partial x}+\hat j\frac{\partial}{\partial y}$, and $H_\rho=\left({\frac{1}{\rho}{\frac{d \rho}{dz}}}\right)^{-1}$ is the density scale height.

This form of the continuity equation suggests two possible flow regimes: for $\frac{\partial \mbox{u}_z}{\partial z} \gg \frac{\mbox{u}_z}{H_{\rho}}$ the motions may be considered nearly divergenceless and isotropic whereas for $\frac{\partial \mbox{u}_z}{\partial z} \ll \frac{\mbox{u}_z}{H_{\rho}}$ the stratification is most important in determining the flow component speeds. Heuristically, small scale overturning eddies would fall in the first regime, while eddies larger than the local scale height would fall in the second, with the largest isotropic eddies increasing in size with depth as the density scale height increases. Thus we expect the dominant balance in Equation~\ref{conteqn} to depend on the length scale of the flow and depth within the convection zone.

Maintaining the mean stratification in a statistically-steady stratified convective flow requires that most of the mass must overturn as the fluid rises through one scale height; over each scale height the density of rising fluid must decrease by a factor of $1/e$, implying that $1-1/e$ of the mass must overturn. Similarly, downwelling fluid must entrain mass at this rate. If the flow geometry is approximated by vertical cylinders of radius $r$ and height $H_{\rho}$, then for all of the mass to overturn within one scale height, $2\pi r H_{\rho}\rho\mbox{u}_h=\pi r^2 \rho\mbox{u}_z$. This yields a characteristic horizontal scale for the motions~\citep{nor09}
\begin{equation}
\label{reqhrho}
r=2\alpha H_{\rho}\frac{\mbox{u}_h}{\vert\mbox{u}_z\vert}\ ,
\end{equation}
\noindent
where $\alpha$ is a factor of order 1 and includes a weak dependence on geometry and the $1/e$ fraction of the mass that does not overturn. We take this length scale to be the crossover between those motions that feel the stratification and those that do not. We demonstrate in the next section that such a crossover is seen in the spectra of three-dimensional simulations. This length scale is also the integral scale of the velocity spectrum in solar-like hydrodynamic simulations~\citep{ste09}, and henceforth we refer to it as the driving or integral scale of the convection.

\subsection{The spectra of horizontal and vertical motions}

Equation \ref{conteqn} can be written as
\begin{equation}
\label{fftconteqn}
i{\bf k}_h\cdot{\tilde{\bf u}}_h = -\frac{\partial\tilde{\mbox{u}}_z}{\partial z}-\frac{\tilde{\mbox{u}}_z}{H_{\rho}}\ ,
\end{equation}
where the overlying tildes indicate the complex Fourier amplitudes resulting from a two-dimensional horizontal Fourier transform at each depth $z$ and ${\bf k}_h$ is the horizontal mode wavevector. By squaring both sides and taking two limits of Equation~\ref{fftconteqn} we can define a relationship between the power in horizontal and vertical motions without directly solving for the phases of the modes. For modes smaller than the integral scale we take the limit $\frac{\partial \tilde{\mbox{u}}_z}{\partial z} \gg \frac{\tilde{\mbox{u}}_z}{H_{\rho}}$, while for larger scale modes we take $\frac{\partial \tilde{\mbox{u}}_z}{\partial z} \ll \frac{\tilde{\mbox{u}}_z}{H_{\rho}}$. Even in these limits, defining the relationship between vertical and horizontal power is difficult for two reasons: when squaring Equation~\ref{fftconteqn}, the cross terms between horizontal modes $\tilde{\mbox{u}}_x$ and $\tilde{\mbox{u}}_y$ on the left side do not have an a priori known form, and the vertical derivative on the right hand side cannot be simply related to the wavenumber of the horizontal Fourier modes.

To proceed we make simplifying assumptions which we have empirically verified to hold in stratified~\citep{lor14} and incompressible turbulence simulations~\citep{min06} as appropriate. At small scales the flow is nearly isotropic and homogeneous, with unstratified homogeneous and isotropic turbulence simulations showing that $k_x^2\tilde{\mbox{u}}_x\tilde{\mbox{u}}_x^* \approx k_y^2\tilde{\mbox{u}}_y\tilde{\mbox{u}}_y^* \approx \frac{\partial \tilde{\mbox{u}}_z}{\partial z}\frac{\partial \tilde{\mbox{u}}_z^*}{\partial z}$, which together with incompressibility yields $\frac{\partial \tilde{\mbox{u}}_z}{\partial z}\frac{\partial \tilde{\mbox{u}}_z^*}{\partial z} \approx \frac{1}{4}k_h^2\tilde{\mbox{\bf{u}}}_h\cdot\tilde{\mbox{\bf{u}}}_h^*$. The simulations also suggest a relationship between the vertical and horizontal gradients, $\frac{\partial \tilde{\mbox{u}}_z}{\partial z}\frac{\partial \tilde{\mbox{u}}_z^*}{\partial z} \approx \frac{1}{4}k_h^2\tilde{\mbox{{u}}}_z\tilde{\mbox{{u}}}_z^*$, and together these yield a relationship between horizontal and vertical power:
\begin{equation}
\label{powersmall}
\tilde{\mbox{\bf u}}_h\cdot\tilde{\mbox{\bf u}}_h^*=\tilde{\mbox{u}}_z\tilde{\mbox{u}}_z^*\ ,
\end{equation}
where $\tilde{\mbox{\bf u}}_h\cdot\tilde{\mbox{\bf u}}_h^*=\tilde{\mbox{u}}_x\tilde{\mbox{u}}_x^*+\tilde{\mbox{u}}_y\tilde{\mbox{u}}_y^*$.
At large scales, the cross terms, which result from squaring the left hand side of Equation~\ref{fftconteqn}, are measured in stratified simulations to be small and are set to zero. This implies that $k_x^2\tilde{\mbox{u}}_x\tilde{\mbox{u}}_x^* + k_y^2\tilde{\mbox{u}}_y\tilde{\mbox{u}}_y^* \approx\tilde{\mbox{u}}_z\tilde{\mbox{u}}_z^*/{H_{\rho}^2}$,
and the horizontal and vertical power in the modes are related as
\begin{equation}
\label{powerlarge}
\tilde{\mbox{\bf u}}_h\cdot\tilde{\mbox{\bf u}}_h^*= \frac{2}{\mbox{k}_h^2H_{\rho}^2}\tilde{\mbox{u}}_z\tilde{\mbox{u}}_z^*\ ,
\end{equation}
where $\mbox{k}_h^2=\mbox{k}_x^2+\mbox{k}_y^2$.

Finally, without approximation, the driving scale that separates these two behaviors (Equation \ref{reqhrho}) can be rewritten as
\begin{equation}
\label{fftreqhrho}
\lambda_h=4\alpha H_{\rho} \frac{\mbox{u}_h}{\vert\mbox{u}_z\vert}\ ,
\end{equation}
where $\lambda_h = {2\pi}/{\mbox{k}_h}$ is the wavelength of the Fourier mode corresponding to a convective cell diameter of $d=2r$. By taking $\alpha$ and $\mbox{u}_h/ \mbox{u}_z\approx 1$ we approximate the driving scale as $\lambda_h\approx4H_\rho$. It is on the basis of these relationships (Equation~\ref{powersmall} at small scales and Equation~\ref{powerlarge} at large scales with the crossover between them given by the driving scale $\lambda_h = 4H_\rho$ at each depth) that we calculate the horizontal velocity power spectrum from the vertical.

\begin{figure}[t]
\plotone{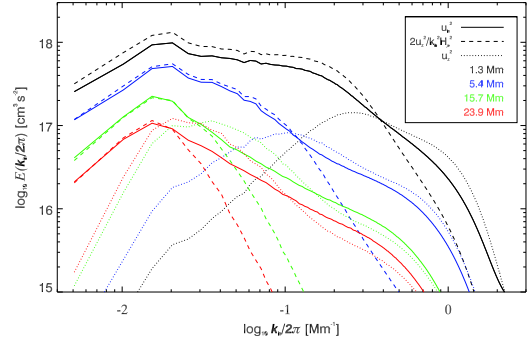}
\caption{\label{fig1} The horizontal velocity power spectra ({\it solid} curves) from hydrodynamic simulations using a Saha equation of state (see \S{3.1} and \S{5.2} for details) at four depths: black 1.3Mm, blue 5.4Mm, green 15.7Mm, and red 23.9Mm below the photosphere. The {\it dashed} and {\it dotted} curves show the horizontal velocity spectra deduced from the vertical velocity based on Equations~\ref{powersmall} and~\ref{powerlarge} respectively. The velocity is averaged over 30 minutes before computing these spectra to remove p-modes. This averaging also reduces power in high wavenumber convective motions.}
\end{figure}

\begin{figure}
\plotone{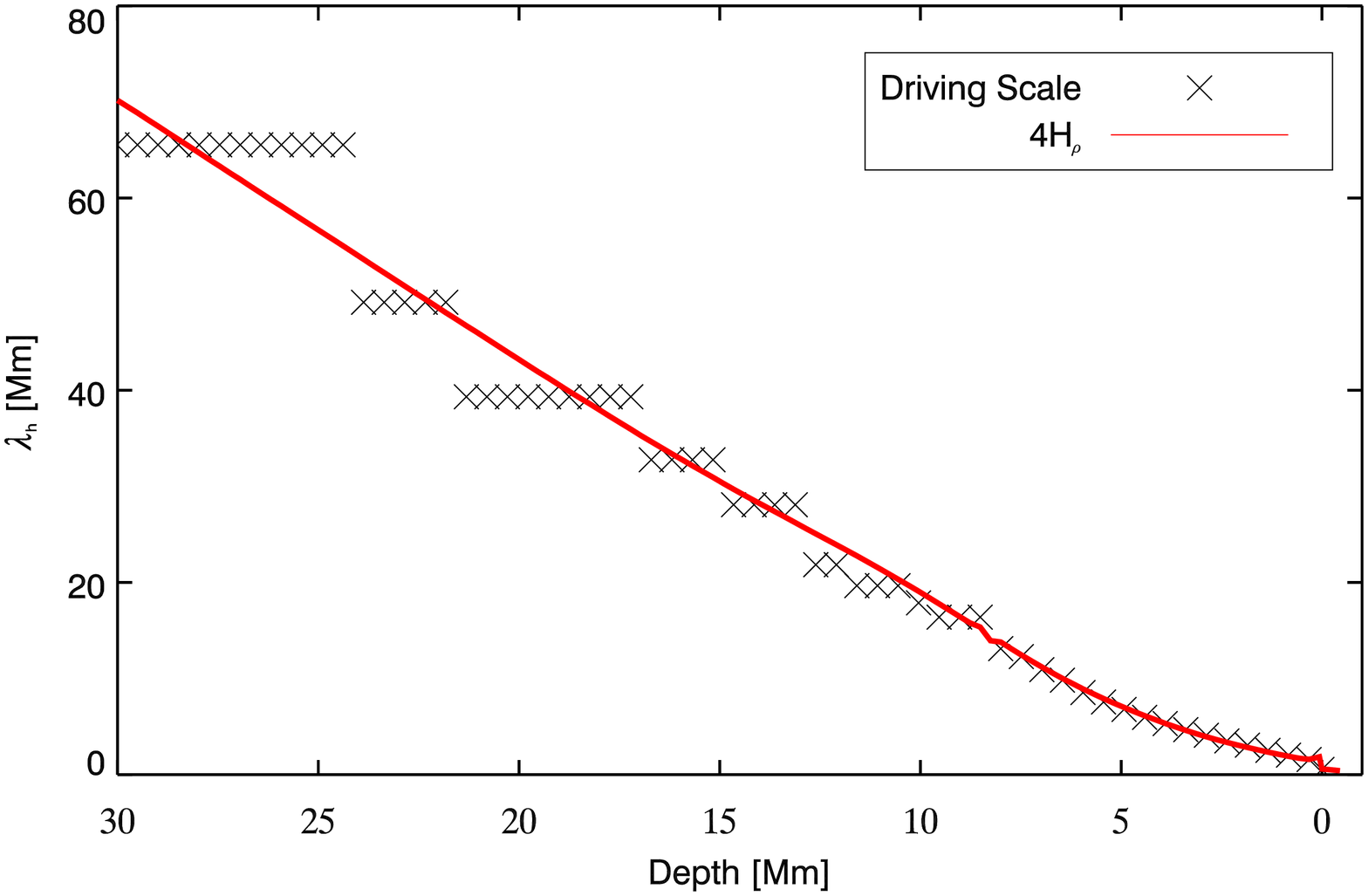}
\caption{\label{fig2} The {\it crosses} show the driving scale in the hydrodynamic simulation. This scale separates the small scale divergenceless motions that follow Equation~\ref{powersmall} and the large scale motions that feel the stratification and follow Equation~\ref{powerlarge}. The driving scale is taken to be the smallest wavenumber where the horizontal velocity spectrum begins to systematically diverge from Equation~\ref{powerlarge}. The {\it solid} (red) curve shows $4H_\rho$ from the hydrodynamic simulation.}
\end{figure}

Analysis of large scale radiative hydrodynamic simulations of solar convection (for details, see \S{5.1} and~\cite{lor14}) helps validate these relationships. Below $1.3$Mm beneath the photosphere, the mass continuity in the simulations matches the anelastic balance (Equations~\ref{conteqn} and \ref{fftconteqn}) to within a few percent. Near the photosphere this balance breaks down because of fluid compressibility, particularly at high wavenumber. We thus restrict our model analysis to depths below 1.3Mm. At low wavenumbers, p-mode contributions can still be important at the shallowest depths. We remove these when comparing the numerical simulations to the model by averaging the simulation velocities over 30 minutes. This averaging also reduces the amplitude of the high wavenumber convective motions, but preserves the relationships between horizontal and vertical flows of Equations~\ref{powersmall} and~\ref{powerlarge}. This is illustrated by Figure~\ref{fig1}, in which the horizontal velocity spectra measured at several depths in a solar-like radiative hydrodynamic simulation are plotted. Overplotted are the horizontal velocity spectrum deduced from the vertical velocity spectra of the simulation at the same depths using Equations~\ref{powersmall} and~\ref{powerlarge} ({\it dotted} and {\it dashed} line-styles respectively). The two component reconstruction of the horizontal velocity spectrum from the vertical reproduces the shape and amplitude of the actual spectrum quite well. Moreover, the driving scale estimate of $4H_\rho$ is in good agreement with the crossover between the two behaviors. Plotted in Figure~\ref{fig2} is the crossover wavenumber as a function of depth (defined as smallest wavenumber in the simulations for which the balance in Equation~\ref{powerlarge} begins to fail, meaning that the difference between the two terms at neighboring larger horizontal wavenumbers is increasingly large). For comparison, $4H_\rho$ is overplotted in {\it red}. They are in good agreement. Note that the discontinuities in the measured values are due to the finite spectral resolution of the simulation; many depths in the simulation appear to have the same crossover scale because there are no modes that can discriminate between them.

\section{Model components}

Having verified the two component continuity balance, we construct the horizontal spectrum of horizontal motions from the horizontal spectrum of the vertical velocity using the relationships derived. To do this we must model the vertical velocity spectrum at each depth. This depends on the driving (integral) scale at that depth, the spectrum of the small scale motions, and the decay rate of the large scale modes that are driven below the height of consideration. We have already defined the driving scale as $4H_\rho$, and we choose the spectrum of the higher wavenumber motions to follow a turbulent cascade with a $k^{-5/3}$ Kolmogorov slope. The Kolmogorov spectrum does not match the spectrum of motions in the hydrodynamic simulations exactly, but we use it in the model for simplicity and in place of an ad hoc fit to the simulations, which themselves may not match the spectrum of solar motions (see \S{4}). The integrated spectral power is determined using the rms velocity of a mixing length model of solar convection (\S{3.1}). Thus the amplitudes of modes with scales larger than the driving scale are determined by their decay with height from the depth at which they were last driven (\S{3.2}), and the remaining power (the rms velocity squared minus the power in large scale modes) is distributed among all modes at the driving scale or smaller according to a Kolmogorov distribution.

\subsection{Mixing length transport by small scale modes}

We employ a simplified hydrostatic mixing length atmosphere of pure hydrogen and helium in Saha equilibrium, integrating the mixing length equations~\citep{pra25,boh58} using the observed density and temperature of the solar photosphere as boundary conditions. We take the convective flux to be equal to the photospheric radiative output ($6.3 \times10^{10} \mbox{ erg} \mbox{ cm}^{-2} \mbox{ s}^{-1}$) as appropriate for efficient convection, and note that this introduces an error in the lower portions of the model where, in the Sun, radiation transports a significant fraction of the energy flux. This error makes a small contribution to the excess model power at the largest scales (\S{4}).

Explicitly, we solve the equation for the convective energy flux $F_c=\frac{1}{2}\rho v C_p T \frac{l}{H_p}(\nabla-\nabla')$ along with that of hydrostatic balance $\frac{dP}{dz}=-\rho g$. Here $\rho$ is the fluid density, $v$ is the velocity, $C_p$ is the specific heat at constant pressure, $T$ is the temperature, $\nabla$ is the mean temperature gradient, $\nabla'$ is the temperature gradient within the convective cell, $P$ is the pressure, and $g=G m(z)/r(z)^2$ is the gravitational acceleration with $r(z)$ the distance from the Sun's center, $m(z)$ the mass within that radius and $G$ the gravitational constant. We employ the equation of state $P=\rho kT/{\mu}$, where $k$ is the Boltzmann constant and $\mu$ is the mean molecular weight of the plasma, and assume that the convective motions are adiabatic, so that $\nabla'=\nabla_{\rm{ad}}=\frac{\partial ln T}{{\partial ln P}}\bigg{\vert}_{\rm{ad}}$, the adiabatic temperature gradient. Finally, the rms convective velocity is given by $v^2=\frac{1}{8}gQ\frac{l^2}{H_p}(\nabla-\nabla')$, where $l$ is the mixing length measured in units of the pressure scale height ${H_P}$. Note that $Q=1-\frac{\partial ln \mu}{\partial ln T}\bigg{\vert}_P$, $C_p$, $\nabla_{\rm{ad}}$, and $\mu$ account for the non-ideal effects of hydrogen and helium ionization (where the number density of each ionization state is determined in collisional equilibrium as a Saha balance). 

The equations are integrated numerically from the photosphere downward, yielding the convective rms velocity and the local density scale height at each depth.

\subsection{Decay of large scale modes with height}

In all simulations of solar convection the amplitude of the vertical velocity at low wavenumbers decreases towards the surface where granular scale convection is dominant. The rate of this decrease for the largest scale convective modes is a fundamental uncertainty in our understanding of solar convection. While global simulations predict giant cell convection throughout much of the convection zone~\citep[e.g.,][]{mie08}, surface observations have only very recently found evidence for weak flows at these scales~\citep{hat13}. Numerical simulations have difficulty directly addressing the photospheric amplitude of large scale motions. They are either of limited extent in depth~\citep{ste09,ust10,lor14} or do not capture the non-ideal and highly compressible nature of the uppermost layers~\citep{mie08}.

\begin{figure}
\plotone{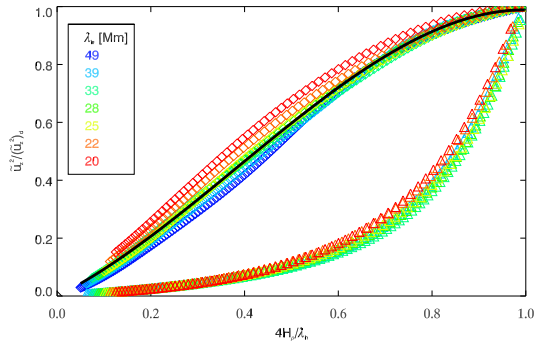}
\caption{\label{fig3} The {\it diamonds} show vertical velocity power from the driving depth, $\lambda_h=4H_\rho$, up to 1.3Mm below the photosphere in the hydrodynamic simulation. The different colors show different modes from 20Mm (red) to 50Mm (purple). The {\it black} curve is the cubic polynomial fit to the observed decay rate. The {\it triangles} show the decay rate for a potential flow (Equation~\ref{expdecay}) for the same range of modes.}
\end{figure}

Because of these uncertainties, we examine two possible vertical velocity amplitude decay profiles. Starting from the depth at which the wavelength of the mode exceeds the integral scale, we decay the modes either by approximating the flow as potential~\citep{avb86} or by using a cubic polynomial fit to the observed decay of modes with wavelengths between 20Mm to 50Mm in the hydrodynamic simulations (shown as {\it diamonds} in Figure~\ref{fig3}).
The potential flow approximation takes the flow to be irrotational, allowing a direct solution to the large-scale continuity balance, written as
\begin{equation}
\frac{\partial\tilde\phi}{\partial z}=-{\mbox{k}}_hH_\rho\tilde\phi\ ,
\end{equation}
where the $\phi$ is the velocity potential with $\tilde{\mbox{u}}_z=\frac{\partial\tilde\phi}{\partial z}$. This yields a profile for the velocity amplitude with height
\begin{equation}
\label{expdecay}
\tilde{\mbox{u}}_z(z)=\tilde{\mbox{u}}_z(z_d) \frac{H_\rho(z)}{H_\rho(z_d)} \exp\left[k_h^2\int_{z_d}^zH_\rho(z')dz'\right] \ ,
\end{equation}
where $z_d$ is the driving depth. This velocity profile can be integrated numerically for any wavenumber $k_h$, the results of which are shown with {\it triangles} in Figure~\ref{fig3}. The polynomial fit, on the other hand, approximates the decay of the modes by a single function determined from the hydrodynamic simulations (solid line in Figure~\ref{fig3}). The fit groups the behavior of all modes between 20 and 50Mm together and is thus inadequate to reproduce the hydrodynamic simulation in detail (see \S{4}). It is employed in the model because of its simplicity. The two schemes are quite different in form, and together provide a test of the sensitivity of the model to this key unknown function.

\subsection{Construction of the model spectrum}

In summary, we construct the model surface horizontal velocity spectrum as follows. To calculate the spectrum of the vertical velocity we: 
\begin{enumerate}
\item
Construct a mixing length model of the solar convection zone integrating from the photosphere downward to 200Mm, the approximate depth of the solar convection zone.
\item
Determine the wavelength of the largest scale mode allowed at the bottom of the model atmosphere, $\lambda_h=4H_\rho$, and use this as the integral (driving) scale (i.e. the lowest wavenumber mode) in a $k^{-5/3}$ turbulent cascade. The highest wavenumber in the spectrum is taken to be the Nyquist frequency of the hydrodynamic simulations on which the model is based ($2\pi/384$km, see $\S{5.1}$), and the spectrum is normalized so that the integrated power is equal to the mixing length velocity squared at the bottom of the model atmosphere.
\item
Move one step up in the atmosphere (a grid spacing of 64km is used to again match the hydrodynamic simulations). Decay modes with wavelengths longer than the local integral scale ($4H_\rho$) using one of the two decay functions discussed in \S{3.2}. Compute the integrated power in the decaying modes and normalize the remaining $k^{-5/3}$ spectrum by the squared mixing length velocity minus the power in the decaying modes.
\item
Repeat Step~3 until the top of the model atmosphere is reached.
\end{enumerate}
\noindent
From the vertical velocity spectrum, the horizontal velocity spectrum at any height is computed using Equations~\ref{powersmall} and \ref{powerlarge}.

\section{Testing the Model}

\begin{figure}
\epsscale{1}
\plotone{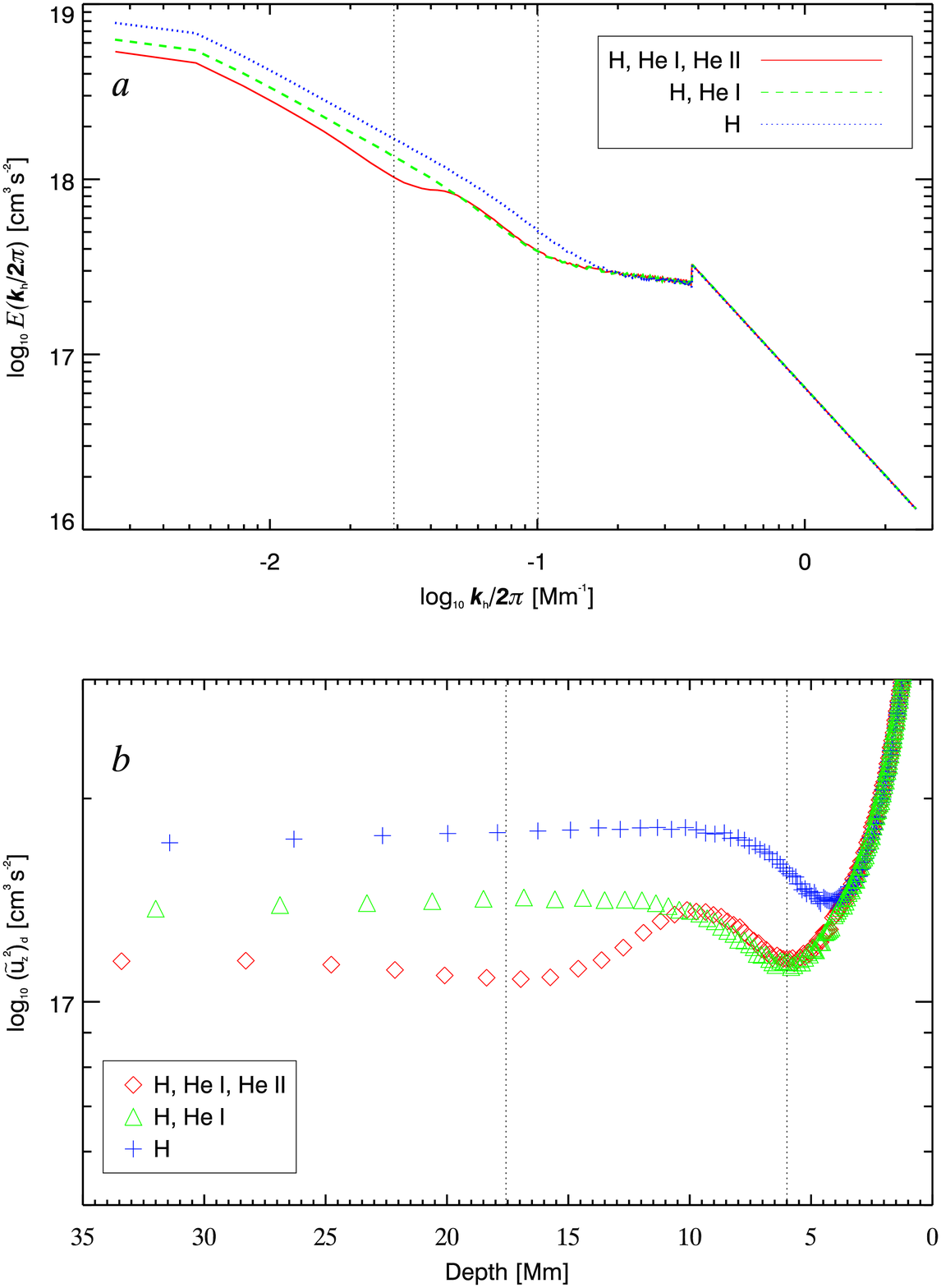}
\caption{\label{fig4}
The spectra computed from the two component model using the decay rate fit to the hydrodynamic simulations. We show the horizontal velocity spectrum at a depth of 1.3 Mm in part {\it a} and the vertical velocity power at the driving depth in part {\it b}. The {\it solid} (red) curve in part {\it a} and {\it diamonds} in part {\it b} (red) shows the spectrum computed from a mixing length atmosphere with a Saha equation of state that includes H, He I, and He II ionization; the {\it dashed} in {\it a} and {\it triangles} in {\it b} (green) is computed from an atmosphere with no He II ionization; and the {\it dotted} in {\it a} and {\it crosses} in {\it b} (blue) is computed from an atmosphere with no He I or II ionization. The vertical {\it dotted} lines show the depths of 50\% He I (6 Mm and $k/2\pi=0.1 \mbox{Mm}^{-1}$) and He II (17.5 Mm and $k/2\pi=0.03 \mbox{Mm}^{-1}$) ionization in part {\it b} and driving scale, where $\lambda_h=4H_\rho$, at those depths in part {\it a}.
}
\end{figure}

We used the model outlined above to compute the horizontal velocity spectrum at a depth of 1.3Mm below the solar photosphere (as previously discussed model assumptions break down above this height and results from hydrodynamic simulations validate this spectra as a close approximation to the surface spectrum for supergranular and larger scale motions). The resulting spectrum is shown as a {\it solid} (red) curve in Figure~\ref{fig4}{\it a}. For clarity we show the spectrum obtained when employing the large scale mode decay rate as measured in the hydrodynamic simulation only (we discuss the potential decay below). The spectrum has two notable low-wavenumber features: monotonically increasing power at scales larger than supergranulation and a small plateau of power at supergranular scales. The monotonic increase of power to lower wavenumbers is not observed on the Sun. The horizontal velocity spectrum of solar motions shows decreasing power at scales larger than supergranulation (see \S5.1). The small supergranular plateau extends from $\lambda_h\sim20-30$Mm (corresponding to $k/2\pi\sim0.03-0.05 \mbox{Mm}^{-1}$ in Figure~\ref{fig4}{\it a}), matching supergranular scales in solar observations. The high wavenumber features of the spectrum, in particular the discontinuity at the 1.3Mm depth integral scale, occur at scales for which the model is ill suited.

To test the sensitivity of the model to the mixing length atmosphere employed we compared the spectrum shown in Figure~\ref{fig4}{\it a} to one computed using a more sophisticated non-local mixing length formulation~\citep{chr96}. The non-local formulation employs the OPAL~\citep{rog92} equation of state and opacities, more carefully accounting for the chemical composition of the convection zone. Importantly, it also accounts for the transport of energy by radiation in the lower portion of the convection zone, which reduces the convective flux and consequent driving amplitudes there. The resulting horizontal velocity spectrum has nearly identical shape as that computed using our simplified Saha balance. Only the largest scale mode shows any notable difference, with the amplitude of that mode somewhat reduced as it is the only mode driven in lower third of the convection zone where the~\cite{chr96} convective velocities are weaker as a result of the more careful accounting of the radiative energy flux.

As there is no a priori expectation for the decay rate of the large scale modes, this aspect of the model is more difficult to assess. We chose to compare the spectrum shown in Figure~\ref{fig4}{\it a} (obtained using the large scale mode decay rate measured in the hydrodynamic simulations) to one employing an analytic potential flow assumption~\citep{avb86} because the later yields an exponential decay of the mode amplitudes (Equation~\ref{expdecay}) and may thus represent a somewhat limiting case. The surface horizontal velocity spectrum computed with the exponential decay shows significant reduction in overall power, particularly at small scales, but quite similar shape at supergranular scales. It exhibits a nearly identical monotonic increase of power at scales larger than supergranulation to that seen in Figure~\ref{fig4}{\it a}. The relative amplitudes of low wavenumber modes is quite insensitive to the imposed mode decay function. The low wavenumber power distribution is primarily determined by the mode amplitudes at depth, with those amplitudes constrained by convective flux requirements of the model atmosphere.

Finally, we looked to validate the model using the three-dimensional hydrodynamic simulations directly. When taking the driving scale ($4H_\rho$) and rms velocity amplitude directly from the simulation itself, rather than from a mixing length atmosphere, the model matches the horizontal velocity spectrum of simulation to within 10\% over the wavenumber band $k/2\pi=0.02-0.15 \mbox{Mm}^{-1}$ ($\lambda_h\sim7-50$Mm) at all depths below 1.3 Mm. This is achieved, however, only by fitting the decay rate of each mode individually and reducing the overall amplitude of the spectrum by a constant offset factor of two. The increased power in the model spectrum results because the power at the driving depth in the model is overestimated by the assumed Kolmogorov power distribution of the isotropic modes. The factor of two can be removed by using a non-Kolmogorov spectrum at depth, but this introduces additional free parameters that can not be constrained by solar observations. This highlights an important result: the shape of the horizontal velocity spectrum in the upper layers of the model is largely determined by the vertical velocity amplitude of the modes at depth. The relative amplitudes of large scale modes in the solar photosphere depends critically on the vertical velocities at the depth. This is further supported by the models inability to reproduce the simulation results for scales $\lambda_h>50$Mm ($k/2\pi<0.02 \mbox{Mm}^{-1}$). For these very large scale motions the driving depth lies near the bottom of the simulation domain and the mode amplitudes, as well as the measured decay rates, are influenced by the simulation lower boundary condition.

\section{Surface Convection Dependence on Motions at Depth}

The model tests discussed above confirm that the monotonically increasing low wavenumber power and much less prominent supergranular plateau are robust features of the horizontal velocity spectrum. That the model can reproduce the shape of the hydrodynamic simulation spectrum validates the underlying assumption that there are two components to the flow separated by the integral (driving) scale which reflects the local scale height at each depth. Larger scale motions are driven deep in the convection zone and decay from below with height. Smaller scale motions behave as isotropic homogenous turbulence. Mismatches between the model and hydrodynamic simulation spectra and observations however raise broader questions. What is the spectrum of solar convective motions at depth and what governs the decay of these motions with height? 

\subsection{The Problem of Excess Low Wavenumber Power}

Both the simplified model and the full three dimensional radiative hydrodynamic spectra show more power than the Sun at scales larger than supergranulation, with that power increasing monotonically toward lower wavenumbers because large scale flows are convectively driven in the deep layers of the domains. It is worth noting that if the solar spectrum matched either the simplified model or the radiative hydrodynamic simulation spectrum, giant cell convection would be relatively easy to observe as the power in these large scale modes would exceed that in supergranulation. 

\begin{figure}
\epsscale{1}
\plotone{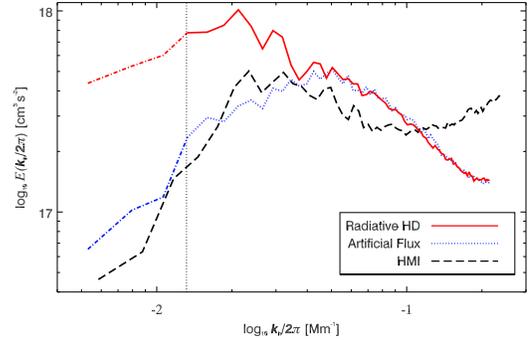}
\caption{\label{fig5} Photospheric velocity power computed from the Coherent Structure Tracking (CST) algorithm of~\citet{rou12} which computes velocities by tracking continuum intensity maps. The {\it dashed} curve is from HMI observations, the {\it solid} (red) curve is from a radiative hydrodynamic simulation in a 196x196x50$\mbox{Mm}^3$ domain using the OPAL equation of state, and the {\it dotted} (blue) curve is from an identical simulation that also includes an artificial energy flux carrying the solar flux below 10 Mm. Modes with $k<0.013\mbox{Mm}^{-1}$ in the simulations (indicated by vertical {\it dotted} fiducial line and {\it dot-dash} linestyle) have driving depths outside simulation domain and consequently have lower photospheric amplitudes than we would expect from a deeper simulation. We use $192\mbox{x}192\mbox{Mm}^2$ HMI images to match simulation domain width and degrade simulation resolution to match the observations ($\sim370$ km). To compute the spectrum of the CST velocities we cut off the two outermost cells and zero-pad by adding twice the number of grid cells in each direction (and multiply all amplitudes by a factor of 4 to maintain integrated spectral power) to remove the influence of the non-periodic boundary.
}
\end{figure}

To make a more direct comparison between our numerical simulations and observations, we employed a Coherent Structure Tracking~\citep[CST,][]{rou12} algorithm to infer the horizontal velocities on large scales from measurements of the motions of granules. In Figure~\ref{fig5} we compare the CST horizontal velocity spectra of a large scale radiative hydrodynamic simulation ({\it solid} red curve) using the OPAL equation of state~\citep{rog92} with that of solar observations ({\it dash} black curve) from the Helioseismic and Magnetic Imager aboard the Solar Dynamics Observatory (HMI/SDO). The measurements in both cases employ a 22 hour sequence of continuum intensity images with each HMI image separated by $45$ seconds and each simulation image separated by $\sim40$ seconds. We break this sequence into 11 two-hour subsets and use the CST method to compute the velocity for each two hour window. The spectrum shown is the average spectrum of those 11 velocity computations. The HMI observations are of 192x192$\mbox{Mm}^2$ region at disk center with a low magnetic activity on 19 June 2010. The simulation solution was computed using the MURaM code~\citep{vog05} in a 196x196x49$\mbox{Mm}^3$ domain with 192x192x64$\mbox{km}^3$ grid spacing~\citep{lor14}.

We note that the spectra in Figure~\ref{fig5} are truncated at high wavenumber because the CST method is not reliable for scales smaller than 2.5Mm. Moreover, low wavenumber modes, those with wavenumbers below $k<0.013\mbox{Mm}^{-1}$ (indicated by the vertical fiducial line and {\it dot-dash} linestyle in Figure~\ref{fig5}) have length scales larger than the integral (driving) scale at the bottom of the simulation domain and consequently have lower photospheric amplitudes than they would likely have in a deeper simulation. Between these extremes are two notable mismatches between the simulation and observation spectra: the simulation shows an excess of power at low wavenumbers and a deficit of power at high wavenumbers when compared to observations. Our very wide and deep simulations resolve supergranular scale motions well but under-resolve granular motions. This leads to an inferred CST velocity with reduced power at high $k$, a result that is inconsistent with the actual simulation velocities and HMI observations. The excess of low wavenumber power is, on the other hand, a fundamental difference between the resolved motions in the simulation and those in observations and is robust, as the CST constrains large scale motions better than small scale motions~\citep{rou12}. Thus understanding the observed solar supergranulation spectrum requires understanding the origin of this low wavenumber reduction in power along with any mechanism that may enhancement power at supergranular scales.

Our simplified mixing length model suggests that the low wavenumber vertical motions are driven deep in the convection zone and decrease in amplitude towards the surface. The radiative hydrodynamic simulations behave similarly (\S{4}), and reducing the vertical flow velocities at depth reduces the low wavenumber horizontal velocity power in the simulated photosphere and improves the match between simulations and observations. We demonstrate this conclusively via simulations in which convective velocities in the deep layers are reduced without changing the mean stratification of the atmosphere (which is also fundamental to the surface spectrum). This was done using an artificial energy transport term. Specifically, we added an artificial flux function to the energy equation that depends only on depth. The artificial flux carries the full solar flux below a specified depth and none of the flux at heights above this. The hyperbolic tangent flux profile employed is 5.12Mm wide centered at 10Mm (where $4H_\rho \sim20\mbox{Mm}$), effectively supporting radiative losses from the photosphere by depositing the heat where the divergence of the function is nonzero. In Figure~\ref{fig5} ({\it dotted} blue curve) we plot the resulting photospheric horizontal velocity spectrum using the same CST method described above. There is substantially reduced power in the photosphere of the artificial flux simulation in those modes that are driven at depths below $\sim10$Mm (scales larger than $\sim20\mbox{Mm}$). This is the region of the domain for which the artificial energy flux is important and consequently convective (rms) velocities are reduced by a factor of $\sim2.5$.

The artificial energy flux experiment confirms the hypothesis that low wavenumber modes are driven deep in the simulated convection zone and imprint as horizontal flows in the surface layers. The photospheric power spectrum reflects a hierarchy of driving scales with depth even in fully nonlinear radiative hydrodynamic simulations. It also suggests that neither the radiative hydrodynamic solutions nor the simplified model spectrum capture the true dynamics of the solar convection below $\sim10$Mm. In other words, in the Sun, low wavenumber flows carry much less of the convective energy flux or transport the energy at substantially lower velocities than expected based on the simulations or the model. Flow/enthalpy correlations, essential to convective transport, may thus not be correctly captured by hydrodynamic simulations. This may be due to their limited resolution or result from the boundary conditions applied. For example, the open boundary condition commonly employed in radiative hydrodynamic simulations of photospheric convection may smooth perturbations in the inflowing plasma. Alternatively, magnetic fields, not included in the simulations we have discussed in this paper, may maintain flow correlations and allow convective transport on smaller scales or at lower velocities than predicted by purely hydrodynamic models. Preliminary results from magnetized simulations favor this hypothesis, though the underlying mechanisms are still under investigation and the effect so far appears insufficient to explain solar observations~\citep{lor14}.

\subsection{Helium Ionization Plays a Minor Role}

The small plateau of power at supergranular scales ({\it solid} red curve in Figure~\ref{fig4}{\it a} at $k_h\sim0.04\mbox{Mm}^{-1}$) reflects the role of helium ionization in determining the convective velocities at depth. Superimposed on the horizontal velocity spectrum in Figure~\ref{fig4}{\it a} we have plotted fiducial vertical lines to highlight the integral (driving) scale at the depths of 50\% He I ($k/2\pi=0.1 \mbox{Mm}^{-1}$, 6 Mm depth) and He II ($k/2\pi=0.03 \mbox{Mm}^{-1}$, 17.5 Mm depth) ionization. The plateau of supergranular power falls between these two fiducial lines. We have also computed the horizontal velocity spectrum for mixing length background atmospheres with an equation of state which does not allow He II or both He I and He II ionization (Figure~\ref{fig4}, {\it dashed} (green) and {\it dotted} (blue) curves respectively). These test atmospheres show a continuous power law increase toward low wavenumbers with no feature at supergranular scales. More precisely these spectra do not show the suppression of power at scales corresponding to the integral scale at the depths of 50\% He I or He II when the ionization processes are disallowed. The differences between the ionizing and non-ionizing spectra at yet lower wavenumbers result because the stratification in the deep layers lies along a different adiabat. The velocity differences at depths below helium ionization, reflected in the low wavenumber horizontal velocity spectra in the near surface, are due to differences in the mean stratification as the medium is nearly fully ionized.  In the region of partial ionization, convective velocities are also influenced by the availability of ionization energy in heat transport via perturbations about the mean ionization state.

Helium ionization is thus responsible for the small supergranular plateau in the model spectrum, albeit in a curious fashion. The ionization of helium yields a slight reduction in the driving scale mode amplitudes in the partially ionized regions (where the driving scale $\lambda_h\sim10$Mm for He I and $\lambda_h\sim35$Mm for He II), producing a small apparent enhancement of power in the upper layers at wavenumbers that lie between them (where $\lambda_h\sim20$Mm). The reduction in mode amplitudes results because ionization energy contributes to the heat transport. In a partially ionized fluid the heat can be transported by ionization state perturbations as well as thermal perturbations~\citep{ras93a,ras93b}, and convective velocities in the mixing length model are thus reduced in partially ionized regions. We note that the mixing length model is a local transport model and does not take into account other effects of ionization such as the increased linear~\citep{ras91} and nonlinear~\citep{ras01} instability of the fluid, though these may play a role in solar convective flows or in our more complete three dimensional simulations. The vertical velocity amplitudes of modes that begin their decay at the depths of helium ionization (the integral or driving scale modes at those depths) are thus suppressed, resulting in a reduction in their horizontal velocity power near the surface. Since ionization energy transport depends on ionization state perturbations, with transport in a fully neutral or fully ionized plasma behaving as an ideal gas, modes with peak amplitudes (those with integral scales equal to $4H_\rho$) at depths that lie between the partially ionized regions (10Mm for example where $\lambda_h\sim20$Mm and $k/2\pi=0.04 \mbox{Mm}^{-1}$) have more power than neighboring modes.

This is illustrated by Figure~\ref{fig4}{\it b}, which shows the vertical velocity power at the driving depth (i.e. the depth where the wavelength of the mode is equal to the integral (driving) scale) for mixing length model atmospheres which do or do not allow He I or He II ionization. The driving scale modes in the regions of partial ionization have lower amplitudes than those outside of it, with minima in the mode amplitudes occurring at the depths of 50\% ionization when it is allowed. The role of hydrogen ionization is difficult to illustrate as its effect is dominant in the surface layers where hydrogen recombination supports radiative losses. The region of H partial ionization is broad in depth and integral to the structure of the radiative boundary layer, and experiments preventing H ionization dramatically alter the mean state of the atmosphere and result in dramatic changes in the velocity spectrum across a wide range of wavenumbers. Thus we do not explicitly consider an atmosphere that disallows hydrogen ionization, but it is clear from Figure~\ref{fig4}{\it b} that the effect of hydrogen ionization on mode amplitudes overlaps that of He I (compare {\it crosses} (blue) and {\it triangles} (green)).

\begin{figure}
\epsscale{1}
\plotone{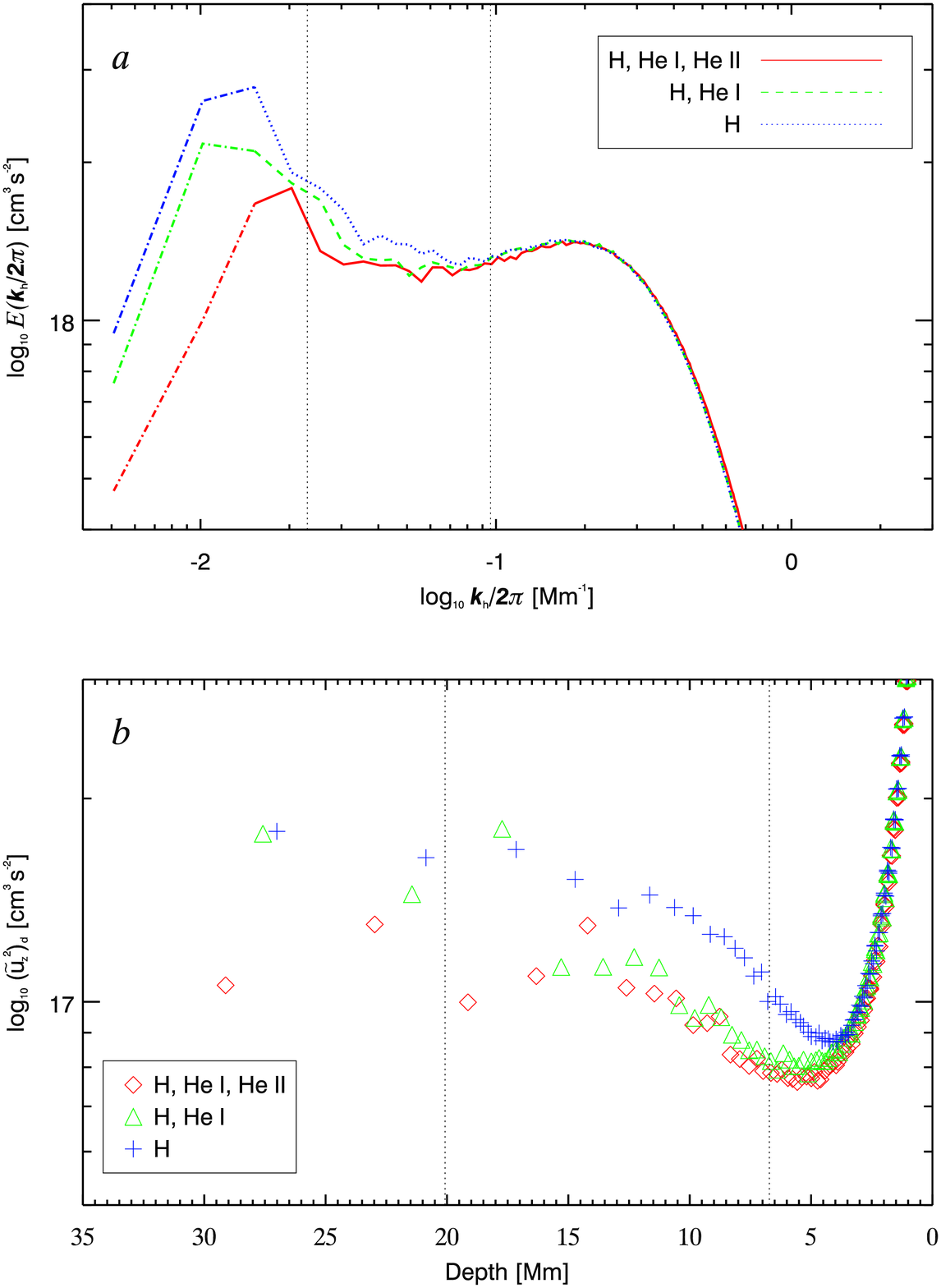}
\caption{\label{fig6}
Horizontal velocity spectrum at the photosphere in part {\it a} and vertical velocity power at the driving depth in part {\it b}. The {\it solid} curve in {\it a} and {\it diamonds} in {\it b} (red) are spectra with a Saha equation of state including H, He I, and He II ionization; the {\it dashed} in {\it a} and {\it triangles} in {\it b} (green) use an equation of state with no He II ionization; and the {\it dotted} in {\it a} and {\it crosses} in {\it b} (blue) use an equation of state with no He I or II ionization. The vertical {\it dotted} lines show the depths of 50\% He I (7Mm and $k/2\pi=0.1 \mbox{Mm}^{-1}$) and He II (20Mm and $k/2\pi=0.025 \mbox{Mm}^{-1}$) ionization in {\it b} and driving scale, where $\lambda_h=4H_\rho$, at those depths in {\it a}. Note that the two largest scale modes are shown in the {\it dot-dash} linestyle here because the driving depth is outside the simulation domain which makes them unreliable.
}
\end{figure}

The simple model we have presented thus suggests that there is an apparent enhancement of photospheric power at $\sim20$Mm scales that occurs because helium ionization reduces the flow speeds in the regions of partial ionization which suppresses power at larger ($\sim35$Mm from He II) and smaller ($\sim10$Mm from He I) scales, not because He II ionization enhances the driving of flows at this depth as has been previously suggested. This apparent enhancement, however, is much smaller than the increased power in observations of solar supergranulation. To investigate the suppression of photospheric power by helium ionization in the context of solar-like convection we use the same Saha equations of state described above in three dimensional radiative hydrodynamic simulations. We use the MURaM~\citep{vog05} code to run simulations that use 192x192$\mbox{km}^2$ horizontal resolution and 64km vertical resolution with 1024x1024x768 grid cells (giving a domain size of 196x196x49$\mbox{Mm}^3$). The results presented here are from more than 5 days of solar time after the simulation has reached a relaxed equilibrium~\citep{lor14}.

Figure~\ref{fig6}{\it a} shows a comparison of the photospheric horizontal velocity spectrum from three such simulations, one which allows H, He I, and He II ionization ({\it solid} red curve), one in which only H and He I ionization are permitted ({\it dashed} green curve), and one with only H ionization ({\it dot} blue curve). The resulting horizontal velocity spectra show similar suppression of photospheric power as that seen the simplified model (Figure~\ref{fig4}) when ionization is allowed. The two lowest wavenumber modes ({\it dot-dash} linestyle) have driving depths outside of the simulation domain and are consequently unreliable and weaker that what would be expected in a deeper simulation. The modes with integral (driving) scales equal to $4H_\rho$ in the regions of partial helium ionization again have reduced amplitudes. This is particularly apparent at the wavenumbers corresponding to modes that peak in the He II partial ionization region ($k/2\pi$ near $0.013 \mbox{Mm}^{-1}$ in Figure~\ref{fig6}{\it a}) which is well separated from the effects of hydrogen ionization. Not allowing He II ionization ({\it dash} green and {\it dot} blue curves) causes a small but significant elevation of photospheric power at those wavenumbers. Disallowing He I ionization ({\it dot} blue curve) induces smaller differences due to the dominant role of hydrogen in the surface layers.

The same reduction in the mode amplitude of the vertical velocity spectrum at the driving depths corresponding to partial helium ionization seen in the simplified model is apparent in these hydrodynamic simulations (Figure~\ref{fig6}{\it b}). Modes with scales equal to $4H_\rho$ at the depths of partial He I and He II ionization have reduced amplitudes, though this reduction is noisier in the simulation than in the mixing length atmosphere (we note that the rms velocity amplitudes, not shown here, also very clearly increase at the nominal ionization depths when ionization is disallowed). This is due to three primary factors: the spectral resolution of the simulation is limited by the domain width, the three dimensional simulation is non-local which makes using a single depth a poor representation of the vertical velocity power that reaches the surface, and the intrinsic temporal variation in the power of the modes below $\sim10$Mm is long compared to the 5 days of simulation time. Moreover, other nonlinear effects of ionization may play some role, as discussed above. These experiments do however confirm, in the context of fully nonlinear three-dimensional radiative hydrodynamic simulations, two important results of the simplified model: the horizontal velocity spectrum in the photosphere reflects the amplitude of the vertical velocity at depth and the reduced amplitude of vertical velocity in the regions of partial helium ionization plays a minor role in shaping the spectrum of supergranular flows at the surface.

\section{Conclusion}

We have constructed a model that computes the horizontal velocity spectrum near the solar surface based on the amplitudes of modes deep in the solar convection zone. The model has three primary features: it is able to match the shape of the photospheric spectrum in three dimensional radiative hydrodynamic simulations, shows a small supergranular scale enhancement of power at 20-30Mm, and an excess of power at lower wavenumbers not seen in observations.

We used the model to examine the role that helium ionization plays in shaping the solar photospheric velocity spectrum. We showed that near the depths of 50\% He I and He II ionization the amplitudes of the vertical motions are reduced because the solar energy flux can be transported at lower velocities due to contribution of ionization energy. This manifests itself as a suppression of horizontal velocity power in the surface layers at scales neighboring supergranulation ($\sim35$Mm scales for He II and $\sim10$Mm scales for He I). We confirmed this effect in three dimensional radiative hydrodynamic simulations that examined convection with and without helium ionization. We conclude that, instead of enhancing a particular flow scale, He I and He II ionization act to highlight supergranular scales by reducing the power in the adjacent modes. This enhancement is, however, smaller in the models than the observed enhancement of solar photospheric power at supergranular scales.

A robust feature of both the model spectrum and the hydrodynamic simulations is an excess of power at low wavenumber when compared to solar observations. This highlights two uncertainties that require further study. First, we do not know the convective flux spectrum in the deep layers of the Sun. While we took the flow spectrum to be Kolmogorov for all scales below the integral scale, this assumption only approximates the spectrum observed in hydrodynamic simulations, and it may significantly underestimate the role small scale motions play in transporting heat through the solar convection zone. Moreover, the large scale hydrodynamics simulations also show excess power at large scales compare to the Sun. Preliminary results from similarly large scale magnetohydrodynamic simulations suggest that magnetic fields may play a role in reducing convective flow speeds or maintaining the correlations required for the energy flux to be carried by smaller scale motions~\citep{lor14}, but as yet these effects are too small to explain observations. Second, we do not know how the amplitude of the vertical motions decreases with height in the solar convection zone. The decay rates (with height) of low wavenumber modes may be influenced by solar rotation and the near surface shear layer which are not included in our analysis. It is likely that the supergranular excess  in the solar power spectrum is largely defined by the observed decrease in power to lower wavenumbers, and has thus been elusive in simulations which show a monotonic increase in power to lower $k$.

The excess low wavenumber power we find in both our simplified model and realistic simulations adds to other recent evidence that large scale flows deep in the solar convection zone are weaker than previously thought. It supports suggestions that numerical simulations more generally may have difficulty matching solar observations if they are required to carry all of the solar energy flux in the resolved modes~\citep{fea14}. Helioseismic observations~\citep{han10,han12} yield estimates of flow velocities that are an order of magnitude or two below those found in either global~\citep[e.g.][]{mie08} or local area~\citep{lor14} simulations. Moreover, as global simulations become more turbulent, with lower diffusivities, flow speeds increase and differential rotation profiles flip to an anti-solar configuration, with a slow equator and fast poles, because rotational constraints are too weak. This transition to anti-solar behavior can be avoided by decreasing the heat flux through the convection zone or increasing the rotation rate~\citep{too13, cha14,hot14}. We found that reducing the convective transport role of large scale modes (by employing an artificial energy flux at all depths below 10 Mm which reduces the deep rms velocities by a factor of $\sim2.5$) can significantly improved the match between the CST spectra of the simulations and observations. These separate lines of evidence all suggest that the Sun transports energy through the convection zone while maintaining very low amplitude large scale motions. Something is missing from our current theoretical understanding of solar convection below $\sim10$Mm.

Acknowledgement: We thank J. Christensen-Dalsgaard for generously providing Model S mixing length velocities. This work was supported in part by NASA award number NNX12AB35G. RHC acknowledges support from DFG SFB 963 "Astrophysical Flow Instabilities and Turbulence" (Project A1).
We acknowledge high-performance computing support from Yellowstone (http://n2t.net/ark:/85065/d7wd3xhc) 
provided by NCAR's Computational and Information Systems Laboratory, sponsored by the National Science Foundation, and from the NASA High-End Computing (HEC) Program through the NASA Advanced Supercomputing (NAS) Division at Ames Research Center.  Computational resources were also provided by NSF-MRI Grant CNS-0821794, MRI-Consortium: Acquisition of a Supercomputer by the Front Range Computing Consortium (FRCC), with additional support from the University of Colorado and NSF sponsorship of NCAR.

\clearpage

\end{document}